\documentclass[aip,amsmath,amssymb,reprint]{revtex4-1}
\usepackage{graphicx}
\usepackage{dcolumn}
\usepackage{bm}
\draft 
\usepackage[utf8]{inputenc}
\usepackage[T1]{fontenc}
\usepackage{mathptmx}
\usepackage{etoolbox}
\usepackage{color}
\begin{document}


\title[Magnon squeezing via reservoir-engineered optomagnomechanics]{Magnon squeezing via reservoir-engineered optomagnomechanics}




\author{Zhi-Yuan Fan}
\author{Huai-Bing Zhu}
\author{Hao-Tian Li}
\author{Jie Li}\thanks{jieli007@zju.edu.cn}
\affiliation{Zhejiang Key Laboratory of Micro-nano Quantum Chips and Quantum Control, School of Physics, and State Key Laboratory for Extreme Photonics and Instrumentation, Zhejiang University, Hangzhou 310027, China}



\begin{abstract}
We show how to prepare magnonic squeezed states in an optomagnomechanical system, in which magnetostriction induced mechanical displacement couples to an optical cavity via radiation pressure. 
We discuss two scenarios depending on whether the magnomechanical coupling is linear or dispersive. We show that in both cases the strong mechanical squeezing obtained via two-tone driving of the optical cavity can be efficiently transferred to the magnon mode. In the linear coupling case, stationary magnon squeezing is achieved; while in the dispersive coupling case, a transient magnonic squeezed state is prepared in a two-step protocol.  The proposed magnonic squeezed states find promising applications in quantum information processing and quantum sensing using magnons.

\end{abstract}

\pacs{}

\maketitle 


\section{Introduction}

The past decade has witnessed the formation and fast development of cavity magnonics, \cite{review1,review2,review3,review4} which studies the interaction between microwave cavity photons with collective spin excitations (magnons) in magnetic materials, such as yttrium iron garnet (YIG).  One of the most prominent advantages of the magnonic system is that it exhibits an excellent ability to coherently couple with a variety of physical systems, including microwave photons, \cite{Huebl2013,Tabuchi2014,Zhang2014} optical photons, \cite{Osada2016,Tang2016,Haigh2016} vibration phonons, \cite{Zhang2016,Li2018,Potts2021,Shen2022} and superconducting qubits,  \cite{Tabuchi2015,DLQ2017,DLQ2020,Xu2023} etc. The magnonic system therefore has great potential for building a hybrid quantum system with multiple components of different nature, which finds broad applications in quantum information science and technology, \cite{review1,review2,review3,review4}  quantum sensing, \cite{Jing2024} and quantum networks.  \cite{LiPRXQ} 

In particular, the coherent couplings among magnons, microwave cavity photons, and vibration phonons form the system of cavity magnomechanics. \cite{CMMreview} In this system, the magnon mode couples to the cavity (mechanical) mode via the magnetic dipole (magnetostrictive) interaction. Theoretical studies indicate that the cavity magnomechanical system offers a promising platform for preparing macroscopic quantum states, such as entangled states \cite{Li2018,Li2019A,Li2021} and squeezed states \cite{Li2019B,WangHF2021,LiNSR} of magnons and phonons, and ground-state cooling of mechanical motion. \cite{Ding20,Jing21,Asjad2023} The magnetostriction induced mechanical displacement can further couple to an optical cavity via radiation pressure, which leads to the formation of the optomagnomechanical (OMM) system. \cite{CMMreview,Fan2023A} The OMM system can act as a foundation for building highly hybrid systems, e.g., by coupling the optical cavity to an ensemble of atoms \cite{Fan2023C}, and the magnon mode to a microwave cavity. \cite{Fan2023D} Relevant entanglement protocols have been put forward, \cite{Fan2023A,Fan2023C,Fan2023D,YRC,DiKe} indicating that the OMM system has many potential applications in quantum information processing and quantum technology.

Here, we propose to prepare magnonic squeezed states using the novel OMM system. We note that several proposals have been offered to prepare magnonic squeezed states. Specifically, they can be achieved by transferring squeezing from microwaves to magnons, \cite{Li2019B} two-tone driving of the magnon mode, \cite{WangHF2021,Qian2024}  two-tone driving of a qubit coupled to the magnon mode, \cite{Guo2023} and exploiting the anisotropy of the ferromagnet, \cite{Kamra2016,Sharma2021} the nonlinearity of the magnetostriction, \cite{LiNSR} and the magnon Kerr effect due to the magnetocrystalline anisotropy, \cite{Asjad2023} etc. In this work, we adopt a different approach by applying the idea of reservoir engineering to the OMM system, and consider two scenarios depending on whether the magnomechanical coupling is linear or dispersive. The strong linear coupling is typically the case where the mechanical frequency is close to the magnon frequency and occurs for higher-frequency phonon modes (in the GHz range) associated with thin YIG films, \cite{An20,Xu21} while the weak dispersive coupling arises for lower-frequency phonon modes (in the MHz range) being much lower than the magnon frequency associated with, e.g., macroscopic YIG spheres. \cite{Zhang2016,Li2018,Potts2021,Shen2022} We first generate strong squeezing of the mechanical mode by driving the optical cavity with a two-tone laser field \cite{}. We then show that in both cases mechanical squeezing can be efficiently transferred to the magnon mode: Stationary magnon squeezing is obtained in the linear coupling case; while transient squeezing is achieved in the dispersive coupling case using a two-step protocol. The quantum squeezing finds many important applications in quantum information processing and quantum sensing in the fields of cavity magnomechanics \cite{Sharma2021,Sun2021,Lu2023,Wan2024,Jing2024,ZhangQF2024,Qian2024} and optomechanics. \cite{He2013,Qiu2020,Lu2024,Tan2020,Jiao2024,LIGO2013,Xu2014,Peano2015,Motazedifard2016,Clark2016,ZhaoW2020}

The article is organized as follows. In Sec.~\ref{linear}, we introduce the OMM system with the linear magnomechanical coupling and provide the Hamiltonian and Langevin equations. We then present the results of the stationary magnon squeezing in Sec.~\ref{resl}. In Sec.~\ref{dispersive}, we describe the OMM system with the dispersive magnomechanical coupling and introduce the two-step operations for preparing a transient magnonic squeezed state. We then present the results and discuss the details related to the experimental realization. Finally, we summarize our findings in Sec.~\ref{conc}. 

\section{The OMM system with linear magnomechanical interaction}
\label{linear}

\begin{figure}[t] 
	\centering
	\includegraphics[width=0.85\linewidth]{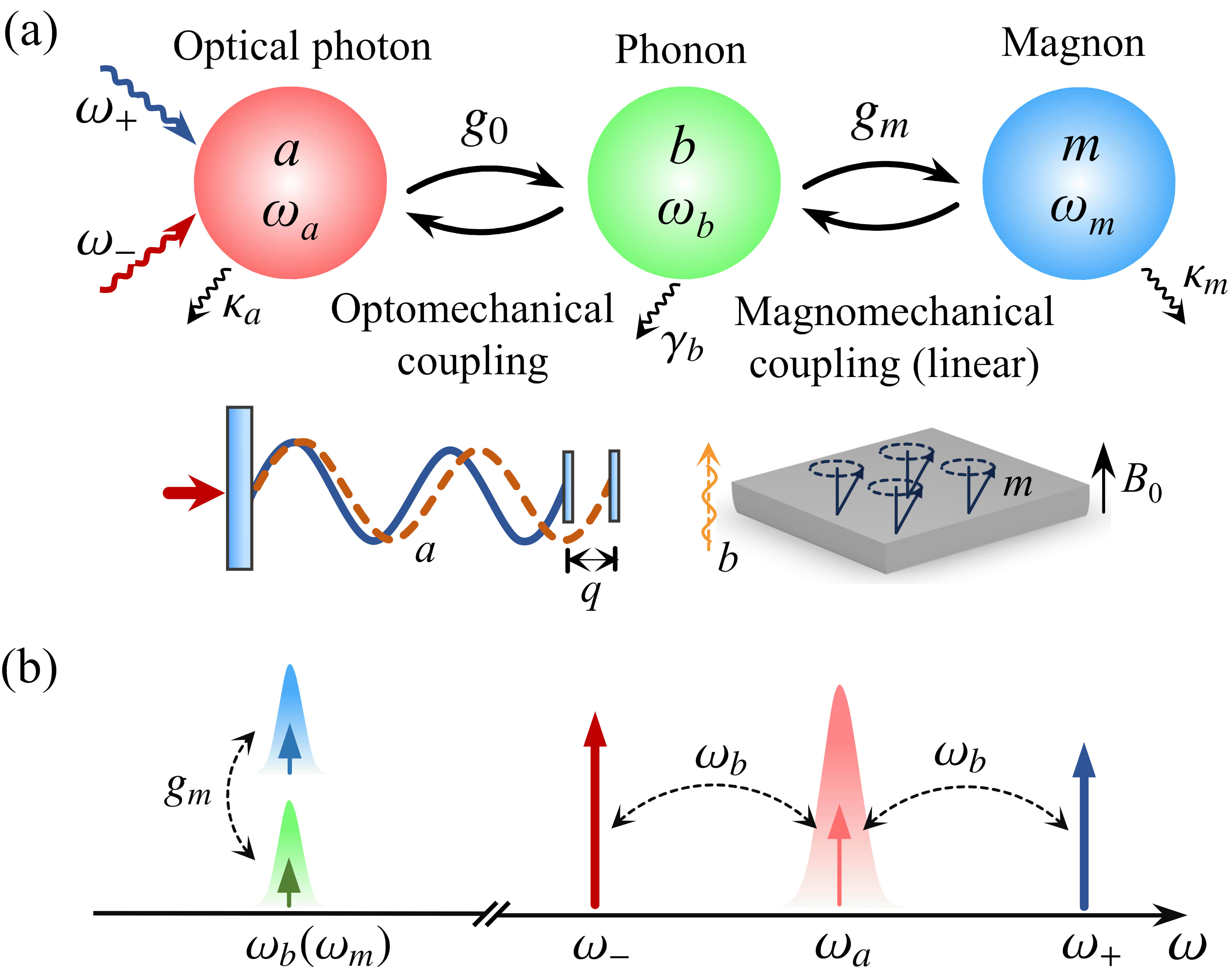}
	\caption{(a) Schematic diagram of the magnon squeezing protocol using the OMM system with a linear magnomechanical coupling.  The mechanical mode ($b$) couples to an optical cavity mode ($a$) via the radiation-pressure interaction and to a magnon mode ($m$) via a linear beam-splitter interaction.  Below are the typical optomechanical system and the magnomechanical system, e.g., a YIG film, with a linear coupling. (b) Mode and drive frequencies used in the protocol. Strong mechanical squeezing is achieved by simultaneously driving the cavity with two laser fields at frequencies $\omega_{\pm}=\omega_a\pm\omega_b$, and the squeezing is further transferred to the magnon mode via the phonon-magnon linear coupling.}
	\label{fig1}
\end{figure}

The OMM system consists of optical cavity photons, ferrimagnetic magnons, and vibration phonons induced by magnetostriction, as depicted in Fig.~\ref{fig1}(a).  Magnons, as quanta of spin wave, are collective excitations of a large number of spins in a ferrimagnet, such as YIG. Originating from the magnetoelastic energy of the magnetic material, magnons can couple to phonons via the magnetostrictive interaction, which describes the interaction between magnetization and elastic strain of the magnetic material. \cite{CMMreview} A linear and strong beam-splitter (BS) type magnon-phonon coupling becomes dominant when the mechanical frequency is high and close to the magnon frequency in the GHz range.  This is typically the case when the magnomechanical sample is a thin YIG film, \cite{An20,Xu21}  as depicted in Fig.~\ref{fig1}(a),  and the strong magnon-phonon coupling leads to magnon polarons. \cite{Kikkawa} The magnomechanical displacement further couples to an optical cavity mode via the radiation pressure or photoelastic effect, \cite{COMreview,Fan2023A}  which can affect the resonance frequency of the cavity field, manifesting a dispersive type interaction. 
Therefore, in the situation of a linear magnon-phonon coupling, the Hamiltonian of the OMM system reads
\begin{equation}
	\label{Hamiltonian}
	\begin{split}
		H/\hbar = &\sum_{o=a,b,m}\omega_o o^\dagger o - g_0 a^\dagger a(b+b^\dagger ) + g_m (m^\dagger b + m b^\dagger) \\
		& + [ ( E_+ e^{-i\omega_+ t} + E_- e^{-i\omega_- t}  ) a^\dagger + \mathrm{H.c.} ],
	\end{split}
\end{equation}
where $o=a$, $b$ and $m$ ($o^\dagger$) are the annihilation (creation) operators of the optical cavity, mechanical, and magnon modes, respectively, and $\omega_o$ are the corresponding resonance frequencies. $g_m$ denotes the coupling strength between the magnon mode and the mechanical mode, which can be strong; $g_0$ is the bare optomechanical coupling strength, which is typically weak, but the effective optomechanical coupling strength $G$ can be significantly enhanced by driving the cavity with a strong laser field. The last term is the driving Hamiltonian corresponding to a two-tone laser field driving the cavity. Such a two-tone driving strategy has been adopted to prepare squeezed states of the mechanical motion in cavity optomechanics  \cite{Tan13,Kronwald2013,Wollman2015,Pirkkalainen2015,Teufel2015}  and cavity magnomechanics. \cite{WangHF2021,Qian2024} Specifically, Fig.~\ref{fig1}(b) shows the frequencies of the driving fields at the Stokes ($\omega_-=\omega_a-\omega_b$) and anti-Stokes ($\omega_+=\omega_a+\omega_b$) scattering sidebands of the optical cavity. Here $E_\pm=\sqrt{\kappa_a P_\pm/\hbar\omega_{\pm}}e^{i\phi_{\pm}}$ represent the coupling strengths between the driving fields and the cavity mode, with $\kappa_a$ being the dissipation rate of the cavity, and $P_\pm$ ($\phi_{\pm}$) the powers (phases) of the driving fields.

Including the dissipation and input noise of each mode, we obtain the following quantum Langevin equations (QLEs) of the system:
\begin{equation}
	\label{QLEs}
	\begin{split}
		\dot{a}=&-i\omega_a a-\frac{\kappa_a}{2}a + ig_0 a(b+b^\dagger ) \\
		&-i(E_+ e^{-i\omega_+ t}+E_- e^{-i\omega_- t}) +\sqrt{\kappa_a} a_{in}, \\
		\dot{b}=&-i\omega_b b - \frac{\gamma_b}{2}b+ig_0 a^\dagger a -i g_m m +\sqrt{\gamma_b} b_{in}, \\
		\dot{m}=& -i\omega_m m-\frac{\kappa_m}{2}m-i g_m b+\sqrt{\kappa_m}m_{in}, 
	\end{split}
\end{equation}
where $\gamma_b$ ($\kappa_m$) is the dissipation rate of the mechanical (magnon) mode. $o_{in}$ ($o=a$, $b$, $m$) are the corresponding input noise operators for the three modes, which are zero-mean and obey the following correlation functions: $\langle o_{in}(t)o_{in}^\dagger(t')  \rangle = [N_o(\omega_o)+1]\delta (t-t')$ and $\langle o_{in}^\dagger(t)o_{in}(t')  \rangle = N_o(\omega_o)\delta (t-t')$. Here, $N_o(\omega_o) = [\mathrm{exp}(\hbar\omega_o/k_B T)-1]^{-1}$ are the mean thermal excitation number of each mode at bath temperature $T$, with $k_B$ the Boltzmann constant.

Since the cavity is strongly driven by the two-tone laser field, which validates the linearization treatment, \cite{Qian2024} we can write each mode operator as the sum of its classical average and quantum fluctuation, i.e., $o=\langle o \rangle + \delta o$ ($o=a$, $b$, $m$). Substituting these decompositions into Eq.~(\ref{QLEs}), we obtain two sets of equations for the classical averages and the quantum fluctuations, respectively. Since the amplitude of the cavity field is dominant at the two driving frequencies $\omega_{\pm}$, we assume that $\langle a\rangle\simeq a_{+}e^{-i\omega_+ t}+ a_{-} e^{-i\omega_- t}$. \cite{Kronwald2013,Qian2024} In typical optomechanical experiments, the cavity frequency shift $2g_0 \mathrm{Re}\langle b\rangle$ due to the radiation pressure is much smaller than the mechanical frequency $\omega_b$. Therefore, the small frequency shift term $ig_0\langle a\rangle(\langle b\rangle+\langle b\rangle^\ast)$ can be safely neglected in solving the set of equations for the classical averages. As a result, we obtain the mean amplitudes 
\begin{equation}
	\label{ave}
	a_{\pm}=\frac{-iE_{\pm}}{i(\omega_{\mp}-\omega_a)+\kappa_a/2} = \frac{-iE_{\pm}}{\kappa_a/2\mp i\omega_b},
\end{equation}
associated with the two frequency components of the cavity field.

The set of linearized equations for the quantum fluctuations of the system, in the interaction picture with respect to $\sum_{o=a,b,m} \hbar\omega_o o^\dagger o$, is given by
\begin{equation}
	\label{fluctuation}
	\begin{split}
		\delta\dot{a}=&-\frac{\kappa_a}{2}\delta a + i(G_- + G_+e^{-2i\omega_b t})\delta b \\
		&+ i(G_+ + G_-e^{2i\omega_b t})\delta b^\dagger +\sqrt{\kappa_a}a_{in}, \\
		\delta\dot{b}=&- \frac{\gamma_b}{2}\delta b -i g_m \delta m+ i(G_-^\ast + G_+^\ast e^{2i\omega_b t})\delta a \\
		&+ i(G_+ + G_- e^{2i\omega_b t})\delta a^\dagger +\sqrt{\gamma_b}b_{in}, \\
		\delta\dot{m}=&- \frac{\kappa_m}{2}\delta m -i g_m \delta b +\sqrt{\kappa_m} m_{in},
	\end{split}
\end{equation}
where $G_{\pm}=g_0 a_{\pm}$ are the effective optomechanical coupling strengths associated with the two driving fields, which are complex in general, but can be set real by adjusting the phases $\phi_{\pm}$ of the driving fields. For simplicity, hereafter we consider real and positive couplings $G_{\pm}$.

Equation (\ref{fluctuation}) is difficult to solve due to the time-dependent oscillating terms. However, under the conditions of $\kappa_{a(m)}$, $\gamma_b$, $G_{\pm} \ll \omega_b$, \cite{Qian2024} we can take the rotating-wave approximation (RWA) and safely neglect those high-frequency oscillating terms. Consequently, we obtain the QLEs in the quadrature form, which can be cast in the following matrix form:
\begin{equation} 
	\label{matrixform}
	\dot{u}(t)=Au(t)+n(t),
\end{equation}
where $u(t)=[\delta X_a(t), \delta Y_a(t), \delta q(t), \delta p(t), \delta X_m(t), \delta Y_m(t)]^T$, with $\delta X_k = (\delta k +\delta k^\dagger)/\sqrt{2}$ and $\delta Y_k = i(\delta k^\dagger -\delta k)/\sqrt{2}$ ($k=a$,$m$) being the fluctuations of the amplitude and phase quadratures of the cavity and magnon modes, respectively, and $\delta q = (\delta b +\delta b^\dagger)/\sqrt{2}$ and $\delta p = i(\delta b^\dagger -\delta b)/\sqrt{2}$ being the fluctuations of the mechanical position and momentum. The drift matrix is given by
\begin{equation}
	A=\begin{pmatrix}
			-\frac{\kappa_a}{2} & 0 & 0 & G_+-G_- & 0 & 0\\
			0 & -\frac{\kappa_a}{2} & G_+ + G_- & 0 & 0 & 0\\
			0 & G_+ - G_- & -\frac{\gamma_b}{2} & 0 & 0 & g_m\\
			G_+ + G_- & 0 & 0 & -\frac{\gamma_b}{2} & -g_m & 0\\
			0 & 0 & 0 & g_m & -\frac{\kappa_m}{2} & 0\\
			0 & 0 & -g_m & 0 & 0 & -\frac{\kappa_m}{2}
	\end{pmatrix},
\end{equation}
and the vector of input noises $n(t)=[\sqrt{\kappa_a} X_{a}^{in}(t), \sqrt{\kappa_a} Y_{a}^{in}(t),$ $ \sqrt{\gamma_b} q^{in}(t), \sqrt{\gamma_b} p^{in}(t),\sqrt{\kappa_m} X_{m}^{in}(t),\sqrt{\kappa_m} Y_{m}^{in}(t)]^T$.

Because of the dynamics of the system is linearized and the input noises are Gaussian, the state of the system at any given time is Gaussian and thereby can be characterized by a $6\times6$ covariance matrix (CM) $V$, with its entries defined as $V_{ij}=\langle u_i(t) u_j(t) + u_j(t) u_i(t)\rangle/2$ ($i,j=1,2,...,6$). The steady-state CM $V$ can be directly achieved by solving the Lyapunov equation \cite{Vitali2007}
\begin{equation}
	\label{Lyaeq}
	AV+V A^T = -D,
\end{equation}
where the diffusion matrix $D=\mathrm{diag}[\kappa_a/2,\kappa_a/2,\gamma_b(N_b(\omega_b)+1/2),\gamma_b(N_b(\omega_b)+1/2),\kappa_m(N_m(\omega_m)+1/2),\kappa_m(N_m(\omega_m)+1/2)]$, which is defined via $D_{ij}\delta(t-t')=\langle n_i(t)n_j(t')+n_j(t')n_i(t) \rangle/2$. Here we assume $N_a(\omega_a)\simeq 0$, due to the high frequency of the optical mode.  

\section{Stationary quantum squeezing of mechanical and magnon modes}
\label{resl}

\begin{figure}[t] 
	\centering
	\includegraphics[width=1\linewidth]{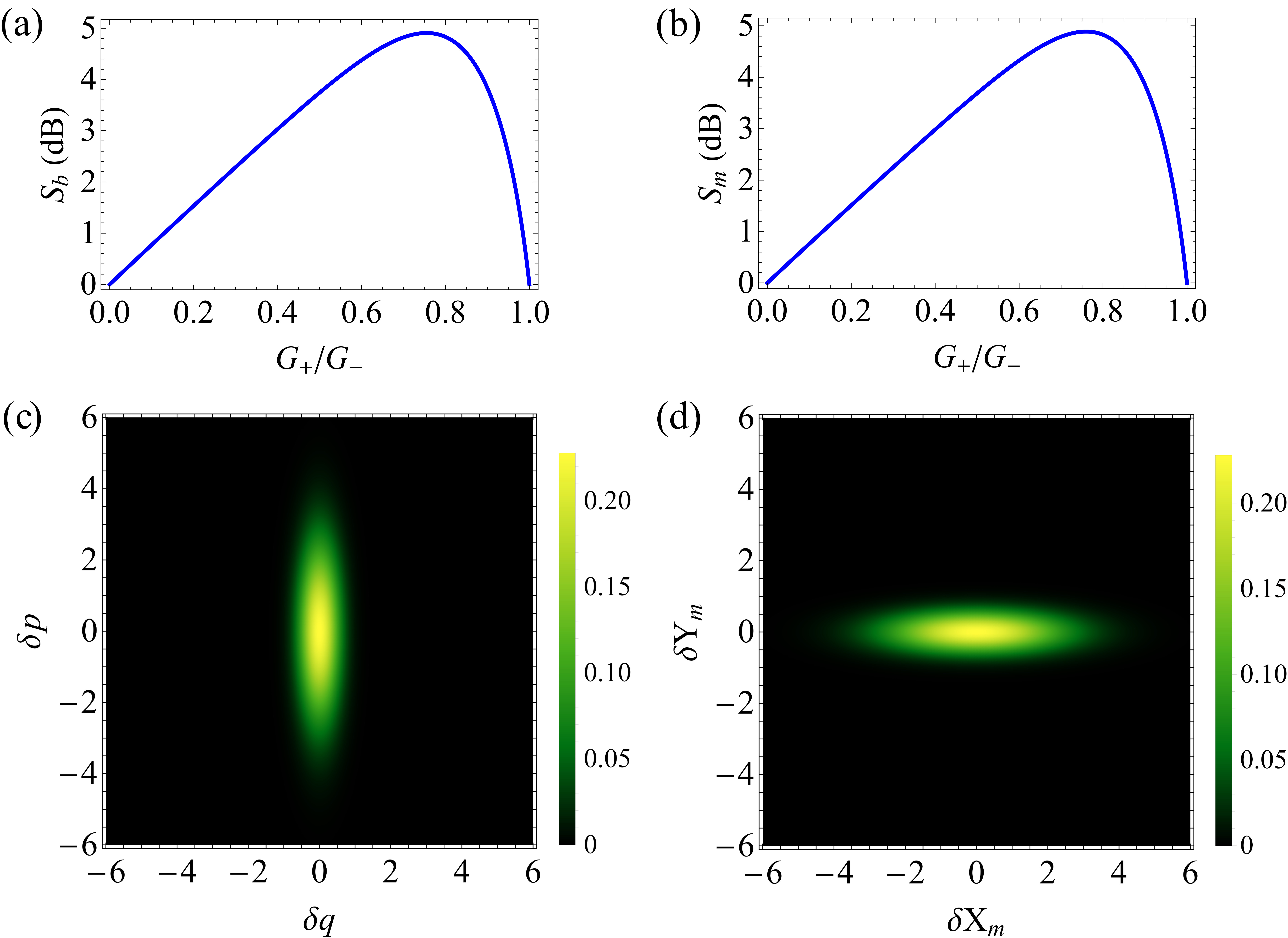}
	\caption{The degree of squeezing of the mechanical displacement (a) and magnon phase quadrature (b) versus $G_+/G_-$. The Wigner function of the squeezed mechanical mode (c) and magnon mode (d), with $G_+/G_- = 0.76$. See text for the other parameters.}
	\label{fig2}
\end{figure}

The physical mechanism of generating quantum squeezing under the two-tone driving field can be explained as follows. Under the RWA, the two driving fields at the blue and red sidebands ($\omega_{\pm}=\omega_a\pm\omega_b$) of the optical cavity can respectively activate the optomechanical parametric down conversion and beam-splitter interactions, and the effective optomechanical interaction can thus be denoted by $\delta a^\dagger (G_+\delta b^\dagger + G_- \delta b)+\mathrm{H.c.}$, cf., Eq.~(\ref{fluctuation}). Using the standard squeezing transformation, \cite{CMMreview,Kronwald2013} we introduce the Bogoliubov mode operator for the mechanical mode, $\delta B=\mathrm{cosh}(r)\delta b+\mathrm{sinh}(r) \delta b^\dagger$, where the squeezing parameter $r$ is defined as $r=\frac{1}{2}\mathrm{ln}\frac{G_-+G_+}{G_--G_+}$. The effective optomechanical interaction is thus rewritten as $\tilde{G} \delta a^\dagger \delta B+\mathrm{H.c.}$, where $\tilde{G}=\sqrt{G_-^2-G_+^2}$ denotes the coupling strength between the cavity mode and the Bogoliubov mode. The above cavity-Bogoliubov beam-splitter interaction can lead to the Bogoliubov mode being significantly cooled to its quantum ground state, which corresponds to the squeezed state of the mechanical mode. By further exploiting the strong phonon-magnon beam-splitter coupling, the squeezing can be efficiently transferred from the mechanical mode to the magnon mode. Consequently, stationary squeezing of the mechanical and magnon modes can be generated under the continuous driving of the two-tone laser field.
 
The degree of squeezing in the dB unit is defined as
\begin{equation}
	S=-10\ \mathrm{log}_{10}[\langle \delta Q(t)^2 \rangle/ \langle \delta Q(t)^2 \rangle_{\mathrm{vac}}],
\end{equation}
where $\langle \delta Q(t)^2 \rangle$ denotes the variance of the quadrature of one specific mode and $\langle \delta Q(t)^2 \rangle_{\mathrm{vac}}=\frac{1}{2}$ corresponds to the vacuum fluctuation.  In Figs.~\ref{fig2}(a) and~\ref{fig2}(b), we show the stationary squeezing of the mechanical displacement and magnon phase quadrature, using the following feasible  parameters: \cite{An20,Xu21,Chan2011,Chan2012,Hong2017} $\omega_m/2\pi=\omega_b/2\pi=10\ \mathrm{GHz}$, cavity resonant wavelength $\lambda=1064\ \mathrm{nm}$, $\kappa_a/2\pi=10^2\ \mathrm{MHz}$, $\kappa_m/2\pi=1\ \mathrm{MHz}$, $\gamma_b/2\pi=10^4\ \mathrm{Hz}$, $g_0/2\pi=10^2\ \mathrm{kHz}$, $g_m/2\pi=10\ \mathrm{MHz}$, $T=10\ \mathrm{mK}$, and a fixed $G_-/2\pi=15\ \mathrm{MHz}$, corresponding to a pump power of $P_-\simeq26.41\ \mathrm{mW}$ for the driving field at frequency $\omega_-=\omega_a-\omega_b$. For the stability reason, we consider $G_+<G_-$. \cite{Qian2024} Clearly, there exists an optimal ratio of $G_+/G_-$ for the maximal  mechanical (magnon) squeezing, as a result of the trade-off between the following two effects: On the one hand, a sufficiently large difference between the two couplings $G_+$ and $G_-$ is required to have a sufficiently large cooling rate $\tilde{G}$ of the Bogoliubov mode; On the other hand, almost equal couplings $G_+ \approx G_-$ is needed to obtain an extremely large squeezing of the mechanical mode. Figures~\ref{fig2}(c) and~\ref{fig2}(d) illustrate the corresponding Wigner function of the squeezed mechanical and magnon modes, with an optimal ratio of $G_+/G_-$.

\begin{figure}[t] 
	\centering
	\includegraphics[width=1\linewidth]{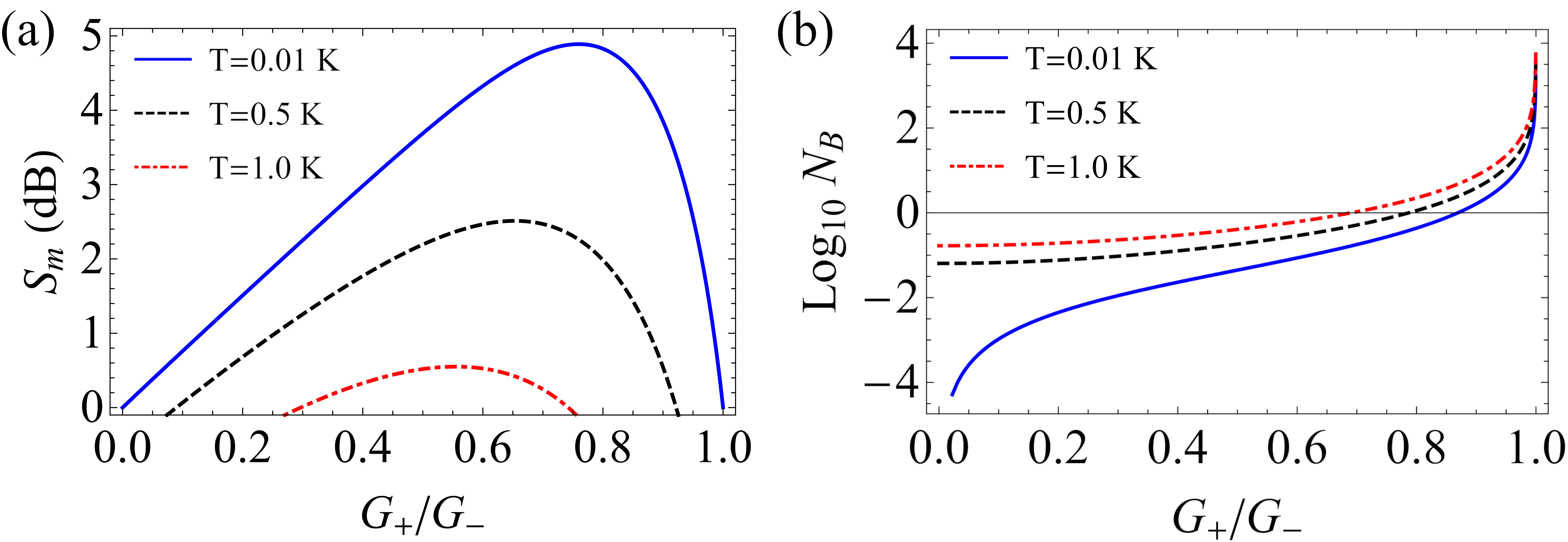}
	\caption{(a) The degree of stationary magnon squeezing $S_m$   and (b) effective phonon number of the Bogoliubov mode $N_B$  versus $G_+/G_-$, at different bath temperatures of $T=0.01$ K (solid), $T=0.5$ K (dashed), and $T=1.0$ K (dot-dahed). The other parameters are the same as in Fig.~\ref{fig2}(b).}
	\label{fig3}
\end{figure}

In Fig.~\ref{fig3}(a), we study the degree of squeezing of the magnon mode $S_m$ versus the ratio $G_+/G_-$ at different bath temperatures. As the temperature rises, the system suffers more thermal noises and a larger cooling rate $\tilde{G}$ is thus needed to efficiently cool the Bogoliubov mode to its ground state, which implies a larger difference between $G_+$ and $G_-$, i.e., a decreasing optimal ratio $G_+/G_-$ (for a fixed $G_-$). This, in turn, reduces the degree of squeezing that can be achieved in the steady state.  Nevertheless, we see that stationary magnon squeezing can be achieved for bath temperature above 1 K, manifesting the robustness of the protocol against thermal noises.  In Fig.~\ref{fig3}(b), we also plot the effective phonon number of the Bogoliubov mode $N_B$ versus the ratio $G_+/G_-$ at different bath temperatures. At the optimal ratio $G_+/G_-$ for each temperature, we obtain $N_B\simeq0.30$ for $T=0.01$ K; 0.39 for $T=0.5$ K; and 0.50 for $T=1.0$ K, indicating that the Bogoliubov mode is cooled (close) to its quantum ground state in all the situations. Note that for $G_+/G_-$ lower than the optimal ratio, the Bogoliubov mode is also in the ground state, but due to a smaller $G_+$ ($G_-$ is fixed), which leads to a smaller squeezing parameter $r$, the degree of squeezing of the mechanical mode (and thus the magnon mode) reduces.

\section{Magnon squeezing with dispersive magnomechanical interaction}
\label{dispersive}

\begin{figure}[t] 
	\centering
	\includegraphics[width=1.0\linewidth]{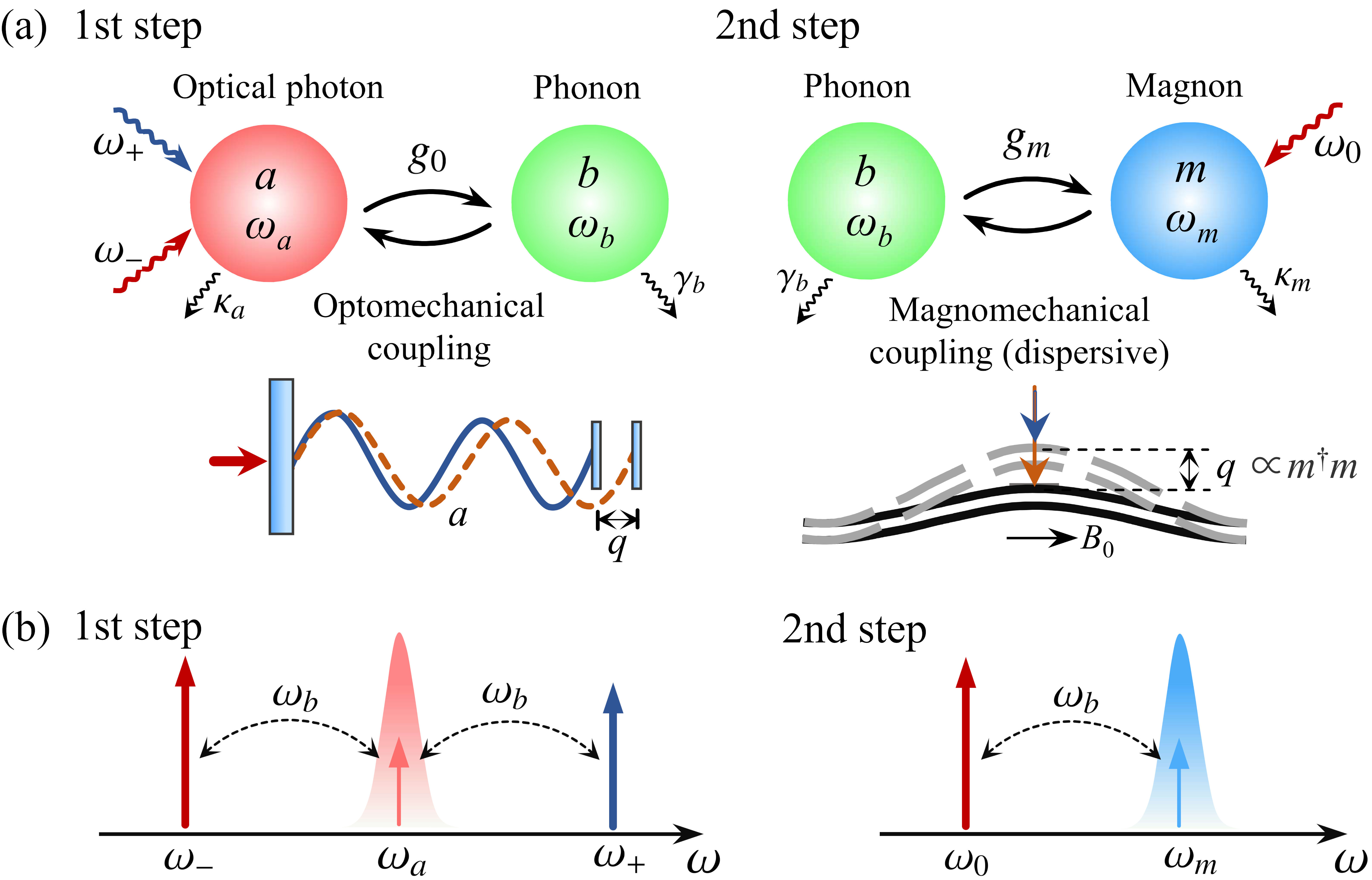}
	\caption{(a) Schematic diagram of the two-step magnon squeezing protocol using the OMM system with a dispersive magnomechanical coupling.  Below are the typical optomechanical system and the magnomechanical system, e.g., a YIG micro bridge, with a dispersive coupling.  (b) Mode and drive frequencies used in the two-step protocol. In the first step, the optical cavity is simultaneously driven by two laser fields at frequencies $\omega_{\pm}=\omega_a\pm\omega_b$ to generate stationary squeezing of the mechanical mode. In the second step, the squeezing is transferred from the mechanical mode to the magnon mode via the magnon-phonon state-swap interaction, realized by driving the magnon mode with a red-detuned microwave field.}
	\label{fig4}
\end{figure}

The above two sections study stationary magnonic squeezed states based on the OMM system with a linear magnomechanical coupling.  However, for lower-frequency mechanical modes, e.g., of a sub-millimeter sized YIG sphere \cite{Zhang2016,Li2018,Potts2021,Shen2022,ShenZ2022} or a micron sized YIG bridge, \cite{Heyroth2019,Fan2023A,Kansanen2021,Arisawa2019} the magnomechanical interaction is dominated by the nonlinear dispersive coupling. \cite{CMMreview}  In this case, the magnetostrictive interaction gives rise to a mechanical displacement of the YIG sample (see Fig.~\ref{fig4}(a)), which is proportional to the magnon excitation number.  The OMM system then has both nonlinear opto- and magnomechanical interactions. The latter weak dispersive coupling hinders an efficient phonon-to-magnon squeezing transfer (which prefers a linear strong coupling as discussed in Secs.~\ref{linear} and~\ref{resl}).  Nevertheless, we show in this section that transient moderate squeezing of the magnon mode can still be achieved by using a two-step protocol, as depicted in Fig.~\ref{fig4}(a).

 
The Hamiltonian of the OMM system with a dispersive magnon-phonon coupling reads
\begin{equation}
	\begin{split}
		H/\hbar =
		 &\sum_{o=a,b,m}\omega_o o^\dagger o - g_0 a^\dagger a(b+b^\dagger )\\ &+ g^{\prime}_m m^\dagger m (b+b^\dagger) + H_{\mathrm{dri}}/\hbar,
	\end{split}
\end{equation}
where $g^{\prime}_m$ is the bare magnomechanical coupling strength, and the effective coupling can be greatly enhanced by driving the magnon mode with a strong microwave field. \cite{Li2018} The last term is the driving Hamiltonian, which is different in the two-step operations.

In the first step, we aim to achieve stationary squeezing of the mechanical mode by driving the optical cavity with a two-tone laser field at frequencies $\omega_{\pm}=\omega_a \pm \omega_b$, as shown in Fig.~\ref{fig4}(b). Due to the weak magnomechanical dispersive coupling and the absence of a microwave drive field to enhance the coupling, the magnon-phonon coupling is very weak. The magnon mode is essentially decoupled from the driven optomechanical system. Therefore, the system under consideration in the first step can be reduced to a two-mode model without the magnon mode. The Hamiltonian then becomes
\begin{equation}
	\begin{split}
		H_{\mathrm{1st}}/\hbar = &\ \omega_a a^\dagger a + \omega_b b^\dagger b - g_0 a^\dagger a (b+b^\dagger) \\
		&+ [ ( E_+ e^{-i\omega_+ t} + E_- e^{-i\omega_- t}  ) a^\dagger + \mathrm{H.c.} ],
	\end{split}
\end{equation}
which gives rise to the following QLEs 
\begin{equation}
	\label{OMqles}
	\begin{split}
		\dot{a}=&-i\omega_a a-\frac{\kappa_a}{2}a + ig_0 a(b+b^\dagger ) \\
		&-i(E_+ e^{-i\omega_+ t}+E_- e^{-i\omega_- t}) +\sqrt{\kappa_a} a_{in}, \\
		\dot{b}=&-i\omega_b b - \frac{\gamma_b}{2}b+ig_0 a^\dagger a +\sqrt{\gamma_b} b_{in}.
	\end{split}
\end{equation}
Following the linearization treatment and taking the RWA as used in Sec. \ref{linear}, we obtain the steady-state averages for the optical cavity $\langle a\rangle\simeq a_{+}e^{-i\omega_+ t}+ a_{-} e^{-i\omega_- t}$ and the mechanical mode $\langle b\rangle\simeq b_{+}e^{-2i\omega_b t} + b_0 + b_{-} e^{2i\omega_b t}$, where
\begin{equation}
	\label{ave1st}
	\begin{split}
		a_{\pm}=&\ \frac{-iE_{\pm}}{\kappa_a/2\mp i\omega_b},\ b_+= \ \frac{i g_0 a_-^\ast a_+}{\gamma_b/2-i\omega_b},\\
		b_0 =&\ \frac{i g_0 (|a_-|^2 + |a_+|^2)}{\gamma_b/2+i\omega_b},\ b_-=\ \frac{i g_0 a_+^\ast a_-}{\gamma_b/2+3i\omega_b}.
	\end{split}
\end{equation}
Similarly, the steady-state CM $V_\mathrm{ab}$ in terms of quantum fluctuations of the optical and mechanical quadratures can be achieved by solving the Lyapunov equation, $A^\prime V_\mathrm{ab}+V_\mathrm{ab} {A^\prime}^T = -D^\prime$, where the drift matrix, in the interaction picture with respect to $\hbar\omega_a a^\dagger a + \hbar\omega_b b^\dagger b$, is given by
\begin{equation}
	A^\prime=\begin{pmatrix}
		-\frac{\kappa_a}{2} & 0 & 0 & G_+-G_- \\
		0 & -\frac{\kappa_a}{2} & G_+ + G_- & 0\\
		0 & G_+ - G_- & -\frac{\gamma_b}{2} & 0 \\
		G_+ + G_- & 0 & 0 & -\frac{\gamma_b}{2}
	\end{pmatrix},
\end{equation}
with $G_{\pm}=g_0 a_{\pm}$, and the diffusion matrix $D^\prime=\mathrm{diag}[\kappa_a/2,\kappa_a/2,\gamma_b(N_b(\omega_b)+1/2),\gamma_b(N_b(\omega_b)+1/2)]$. 

\begin{figure}[t] 
	\centering
	\includegraphics[width=0.63\linewidth]{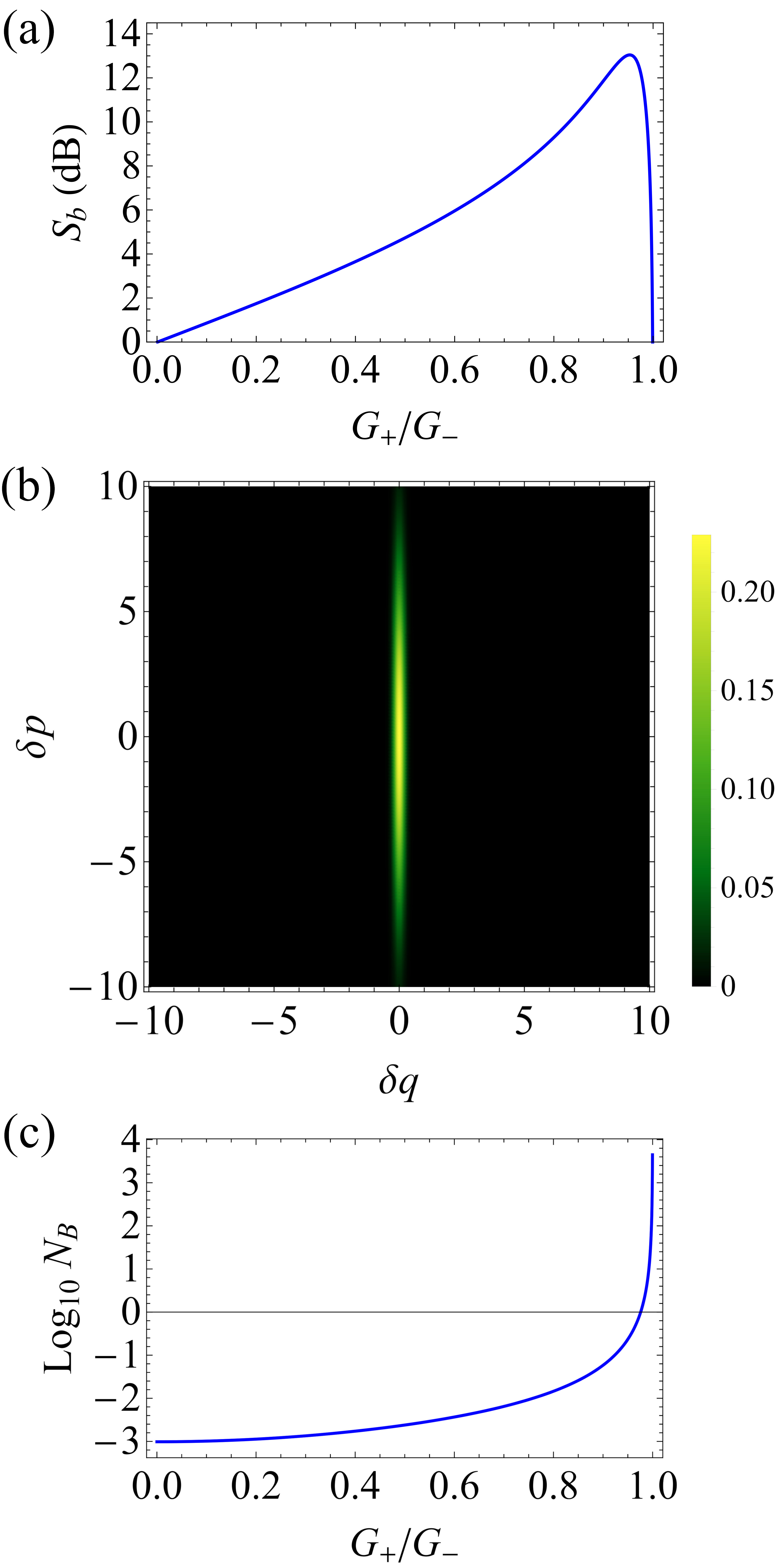}
	\caption{(a) The degree of squeezing $S_b$ of the mechanical mode versus $G_+/G_-$. (b) The Wigner function of the squeezed mechanical mode, with $G_+/G_-=0.95$.  (c) The effective phonon number of the Bogoliubov mode versus $G_+/G_-$. See text for the other parameters.}
	\label{fig5}
\end{figure}

In Fig.~\ref{fig5}(a), we plot the degree of squeezing $S_b$ of the mechanical mode versus $G_+/G_-$, using experimentally feasible  parameters: \cite{Arcizet2006,Favero2007,Simon2009b}  $\omega_b/2\pi=10^2\ \mathrm{MHz}$, cavity resonant wavelength $\lambda=1064\ \mathrm{nm}$, $\kappa_a/2\pi=2\ \mathrm{MHz}$, $\gamma_b/2\pi=10^2\ \mathrm{Hz}$, $g_0/2\pi=1\ \mathrm{kHz}$, and $T=10\ \mathrm{mK}$.  We take $G_-/2\pi=0.3\ \mathrm{MHz}$ corresponding to a power $P_- \simeq 0.53$ mW of the driving field at frequency $\omega_-=\omega_a-\omega_b$. A strong squeezing (13 dB) of the mechanical displacement is achieved at the optimal ratio $G_+/G_-\simeq 0.95$, corresponding to a power $P_+ \simeq 0.48$~mW of the driving field at frequency $\omega_+=\omega_a+\omega_b$.  Figure~\ref{fig5}(b) shows the corresponding Wigner function of the squeezed mechanical mode.  Similarly, we plot the effective phonon number of the Bogoliubov mode versus $G_+/G_-$ in Fig.~\ref{fig5}(c). The Bogoliubov mode is also cooled to the ground state with $N_B\simeq 0.23$ at the optimal ratio $G_+/G_-\simeq 0.95$.

When the stationary and strong mechanical squeezing is achieved in the first step, we then switch off the two driving fields. After a short period, in the time scale of $\kappa_a^{-1}\ll \tau_0 \ll \gamma_b^{-1}$, when the cavity photons completely dissipate, while the mechanical state remains almost unchanged, we then apply a red-detuned microwave field to drive the magnon mode (Fig.~\ref{fig4}(b)) to activate the magnomechanical anti-Stokes scattering, which realizes the magnon-phonon state-swap interaction. \cite{Li2019B}  This subsequent state-swap operation can transfer the squeezing from the mechanical mode to the magnon mode. Note that in the second step, the system can also reduce to a two-mode model, because the cavity photons already died out before applying the magnon drive. The optical cavity is essentially decoupled from the magnomechanical system, and the Hamiltonian of the magnomechanical system is given by
\begin{equation}
	\begin{split}
		H_{\mathrm{2nd}}/\hbar =&\ \omega_m m^\dagger m + \omega_b b^\dagger b + g^{\prime}_m m^\dagger m (b+b^\dagger) \\
		&+ ( \Omega m^\dagger e^{-i\omega_0 t} + \mathrm{H.c.}).
	\end{split}
\end{equation}
The last term is driving Hamiltonian, where $\Omega=\frac{\sqrt{5}}{4}\gamma\sqrt{N}B_d e^{i\phi_0}$ is the Rabi frequency, characterizing the coupling strength between the magnon mode and the driving magnetic field with amplitude $B_d$, frequency $\omega_0$, and phase $\phi_0$. \cite{Li2018} Here $\gamma$ is the gyromagnetic ratio and $N$ is the total number of spins in the YIG crystal.

Since the two-tone laser drive is switched off (the mechanism of generating persistent squeezing is absent), the mechanical squeezing gradually reduces due to its interaction with the thermal bath. Therefore, in the second step the magnon mode will not be prepared in a stationary squeezed state, but we shall implement a {\it dynamical} squeezing transfer from mechanics to magnons.  This is realized by driving the magnon mode with a microwave field with limited duration, and the drive is switched off when the magnon mode evolves into a maximally squeezed state. 
Under the driving field, the QLEs in the interaction picture with respect to $\hbar\omega_0 m^\dagger m$ are given by
\begin{equation}
	\label{QLEs2}
	\begin{split}
		\dot{m} = & -i\Delta_m m-\frac{\kappa_m}{2}m -i g^{\prime}_m m(b+b^\dagger)-i\Omega+\sqrt{\kappa_m}m_{in}, \\
		\dot{b} = & -i\omega_b b-\frac{\gamma_b}{2}b -i g^{\prime}_m m^\dagger m +\sqrt{\gamma_b}b_{in},
	\end{split}
\end{equation}
where the detuning $\Delta_m=\omega_m-\omega_0$ is set to be $\Delta_m=\omega_b$ (Fig.~\ref{fig4}(b)) to activate the magnomechanical state-swap interaction.

Similarly, the above QLEs can be separated into two sets of equations, which describe the time evolution of the averages and fluctuations, respectively. Specifically, the dynamical solutions of the averages $\langle m\rangle_t$ and $\langle b\rangle_t$ can be obtained by numerically solving the equations
\begin{equation}
	\label{mean}
	\begin{split}
		\dot{\langle m\rangle}_t=&-i\Delta_m\langle m\rangle_t -\frac{\kappa_m}{2} \langle m \rangle_t  -i g_m \langle m \rangle_t (\langle b \rangle_t + \langle b \rangle_t^\ast )-i\Omega, \\
		\dot{\langle b\rangle}_t=&-i\omega_b\langle b\rangle_t -\frac{\gamma_b}{2} \langle b \rangle_t  -i g_m \langle m \rangle_t^\ast \langle m \rangle_t,
	\end{split}
\end{equation}
with initial conditions $\langle m\rangle_{t=0}$ and $\langle b\rangle_{t=0}$. The dynamical solutions of the averages are necessary because they will affect the dynamics of the quantum fluctuations, e.g., the effective coupling strength $G_{m_t}=g^\prime_m \langle m\rangle_t$ in the linearized equations for quantum fluctuations is determined by $\langle m\rangle_{t}$.

Moving to another interaction picture by introducing the slowly moving operators with tildes: $\delta \tilde {m}=\delta m e^{i\tilde{\Delta}_mt }$ and $\delta \tilde {b}=\delta b e^{i\omega_b t }$, \cite{Li2021} the QLEs for the quantum fluctuations are given by (with the tildes being dropped for simplicity)
 \begin{equation}
 	\label{fluctuation2}
 	\begin{split}
 		\delta\dot{m}=&-\frac{\kappa_m}{2}\delta m -i G_{m_t}(\delta b+\delta b^\dagger e^{2i\omega_b t})+ \sqrt{\kappa_m}m_{in},\\
 		\delta\dot{b}=&-\frac{\gamma_b}{2}\delta b -i(G_{m_t}^\ast\delta m + G_{m_t}\delta m^\dagger e^{2i\omega_b t}) + \sqrt{\gamma_b}b_{in},
 	\end{split}
 \end{equation}
where $\tilde{\Delta}_m = \Delta_m+g^{\prime}_m (\langle b \rangle_t + \langle b \rangle_t^\ast )\simeq \omega_b$ is the effective detuning and $G_{m_t}=g^\prime_m \langle m\rangle_t$ is the time-dependent effective coupling strength. Under the conditions of $\kappa_m$, $\gamma_b$, $G_{m_t} \ll \omega_b$, we can take the RWA and neglect the nonresonant oscillating terms in Eq.~(\ref{fluctuation2}), and consequently obtain
 \begin{equation}
	\label{fluctuation2RWA}
	\begin{split}
		\delta\dot{m}=&-\frac{\kappa_m}{2}\delta m -i G_{m_t}\delta b+ \sqrt{\kappa_m}m_{in},\\
		\delta\dot{b}=&-\frac{\gamma_b}{2}\delta b -iG_{m_t}^\ast\delta m + \sqrt{\gamma_b}b_{in},
	\end{split}
\end{equation}
which clearly reveal a magnomechanical beam-splitter interaction, and it enables the transfer of the squeezing from mechanics to magnons.  The above equations can be written in the quadrature form, and cast in the following matrix form:
\begin{equation}
	\label{dynequ}
	\dot{u}_{\mathrm{mb}}(t)=A^{\prime\prime}(t) u_{\mathrm{mb}}(t)+n^{\prime\prime}(t),
\end{equation}
where $u_{\mathrm{mb}}(t)=[\delta X_m(t), \delta Y_m(t), \delta q(t), \delta p(t)]^T$ is the vector of quadrature fluctuations, $n^{\prime\prime}(t)=[\sqrt{\kappa_m} X_{m}^{in}(t), \sqrt{\kappa_m} Y_{m}^{in}(t),$ $\sqrt{\gamma_b} q^{in}(t),\sqrt{\gamma_b} p^{in}(t)]^T$ is the vector of input noises, and 
\begin{equation}
	A^{\prime\prime}(t)=\begin{pmatrix}
		-\frac{\kappa_m}{2} & 0 & \mathrm{Im}G_{m_t} & \mathrm{Re}G_{m_t}\\
		0 & -\frac{\kappa_m}{2} & -\mathrm{Re}G_{m_t} & \mathrm{Im}G_{m_t}\\
		-\mathrm{Im}G_{m_t} & \mathrm{Re}G_{m_t} & -\frac{\gamma_b}{2} & 0\\
		-\mathrm{Re}G_{m_t} & -\mathrm{Im}G_{m_t} & 0 & -\frac{\gamma_b}{2}
	\end{pmatrix}
\end{equation}
is the time-dependent drift matrix. The squeezing of the magnon mode can be evaluated by solving the dynamical CM $V_{\mathrm{mb}}(t)$ of the magnomechanical system, via numerically solving
\begin{equation}
	\label{evolution}
	\dot{V}_{\mathrm{mb}}(t)=A^{\prime\prime}(t)V_{\mathrm{mb}}(t)+V_{\mathrm{mb}}(t) A^{\prime\prime}(t)^{T}+D^{\prime\prime},
\end{equation}
with the initial CM $V_{\mathrm{mb}}(0)$ at the beginning of the second step (the time the magnon drive is applied), where $D^{\prime\prime}=\mathrm{diag}[\kappa_m(N_m(\omega_m)+1/2),\kappa_m(N_m(\omega_m)+1/2),\gamma_b(N_b(\omega_b)+1/2),\gamma_b(N_b(\omega_b)+1/2)]$ is the diffusion matrix. To check the validity of the RWA, we also calculate $V_{\mathrm{mb}}(t)$ by numerically solving Eq.~(\ref{fluctuation2}) without taking the RWA for comparison.
\begin{figure}[t] 
	\centering
	\includegraphics[width=1\linewidth]{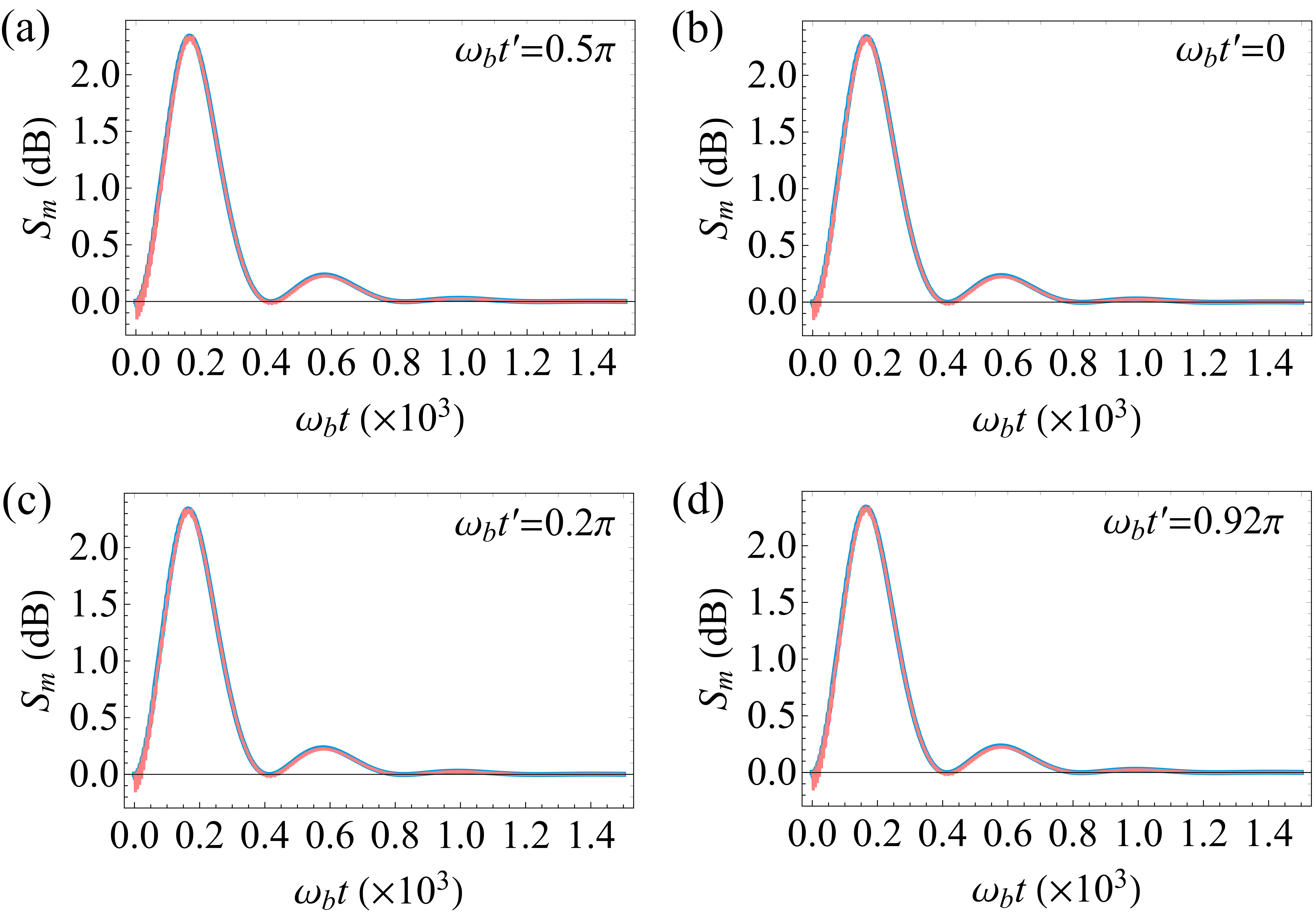}
	\caption{Time evolution of the degree of magnon squeezing $S_m$ in the second step for different switch-off times of the two-tone drive: (a) $\omega_bt^\prime=0.5\pi$, (b) $\omega_bt^\prime=0$, (c) $\omega_bt^\prime=0.2\pi$, and (d) $\omega_bt^\prime=0.92\pi$, with the RWA (blue curves) and without the RWA (pink curves). See text for the other parameters.}
	\label{fig6}
\end{figure}

Figure~\ref{fig6} shows the time evolution of the variance of the magnon phase quadrature $\langle \delta Y_m(t)^2\rangle$ starting from the time when the magnon drive is applied. We adopt experimentally feasible  parameters: \cite{Fan2023A,Heyroth2019,Kansanen2021,Arisawa2019}  $\omega_m/2\pi=10$ GHz, $\omega_b/2\pi=10^2$ MHz, $\kappa_m/2\pi=1$ MHz, $\gamma_b/2\pi=10^2$ Hz, $g^\prime_m=10$ Hz, $\Delta_m= \omega_b$, $T=10$ mK, and $P\simeq 4.93$ mW for a $10\times3.0\times1.5\ \mu m^3$ YIG micro bridge. Figure~\ref{fig6} is obtained by numerically solving Eq.~(\ref{mean}) with the initial conditions $\langle m\rangle_{t=0}=0$ and $\langle b\rangle_{t=0}$ and solving Eq.~(\ref{evolution}) with the initial $V_{\mathrm{mb}}(0)=[V_m^{\mathrm{th}},0;\ 0,\ V_b^{\mathrm{opt}}]$.
Here, $V_m^{\mathrm{th}}=\mathrm{diag}[N_m(\omega_m)+\frac{1}{2},N_m(\omega_m)+\frac{1}{2}]$ denotes that the magnon mode is initially in a thermal state with the mean thermal excitation number $N_m$, and $V_b^{\mathrm{opt}}$ is the CM of the squeezed mechanical mode prepared in the first step (cf., Fig.~\ref{fig5}(b)), which remains almost unchanged in the free evolution process with duration of $\tau_0 \ll \gamma_b^{-1}$. Specifically, we consider $\kappa_a^{-1}  \ll \tau_0=\frac{4\pi}{\kappa_a}\ll \gamma_b^{-1}$, which is long enough for cavity photons to completely decay. 

We note that the initial condition $\langle b\rangle_{t=0}$ used for solving Eq.~(\ref{mean}) depends on the time the two-tone drive is switched off. This can be seen from the results of Eq.~(\ref{ave1st}), where the steady-state averages of the cavity and mechanical modes are periodically oscillating, a typical feature with a two-tone drive. Therefore, different turn-off times (of the two-tone drive) give different initial conditions for solving the equations governing the free evolution process with duration $\tau_0$, i.e., the QLEs (\ref{OMqles}) without the driving terms. This further leads to different initial conditions $\langle b\rangle_{t=0}$ for solving Eq.~(\ref{mean}) in the second step. To study the influence of the switch-off time, we show in Fig.~\ref{fig6} the magnon squeezing for four different turn-off times within a mechanical period $\omega_bt^\prime=0$, $0.2\pi$, $0.5\pi$ and $0.92\pi$ after the mechanical mode reaches the steady state under the two-tone drive. Clearly, the dynamical evolution of the magnon squeezing is almost unaffected. To verify the validation of the RWA used for plotting Fig.~\ref{fig6} (blue curves), we also compare the results with those obtained without taking the RWA (pink curves), i.e., by solving Eq.~(\ref{fluctuation2}). Obviously, the RWA is a very good approximation under the parameters of Fig.~\ref{fig6}.

Figure~\ref{fig6} indicates that a dynamical magnon mode squeezing of $S_m^{\rm max}=2.3$ dB can be achieved in the second step. The microwave driving field can be turned off when the squeezing reaches its maximum, and the transient magnon squeezing can be further transferred to a microwave field using their state-swap interaction \cite{Li2019B}, which is useful for quantum information processing or quantum metrology tasks.

\section{Conclusions}
\label{conc}

We have presented two protocols for preparing magnonic squeezed states in two configurations of an OMM system by exploiting reservoir engineering. In the linear (dispersive) magnomechanical coupling case, stationary (transient) magnonic squeezed states can be generated. The magnon squeezing is transferred from the squeezed mechanical mode that is produced via two-tone driving of the optical cavity. The phonon-magnon state-swap interaction is responsible for the squeezing transfer, which is an intrinsic interaction in the linear coupling case and can be effectively activated in the dispersive coupling case by applying a red-detuned magnon drive. The magnonic squeezed state of a large-size YIG sphere is a macroscopic quantum state and thus useful in testing the limits of quantum mechanics. The magnonic squeezed states also find broad applications in quantum information processing, quantum metrology, and quantum sensing based on magnons. 


\section*{Acknowledgements}
We thank J. Cheng and H. Qian for useful discussions. This work has been supported by National Key Research and Development Program of China (Grant No. 2022YFA1405200) and National Natural Science Foundation of China (Grant No. 92265202).

\section*{Data Availability}
The data that support the findings of this study are available from the corresponding author upon reasonable request.

\end{document}